\documentclass[prl,showpacs,superscriptaddress,twocolumn]{revtex4}
\usepackage{dcolumn}
\usepackage{graphicx}
\begin{document}

\raggedbottom

\title{Multiple packets of neutral molecules revolving for over a mile}

\author{Peter C. Zieger}
\affiliation{Fritz-Haber-Institut der Max-Planck-Gesellschaft,
Faradayweg 4-6, 14195 Berlin, Germany}

\author{Sebastiaan Y.T. van de Meerakker}
\affiliation{Fritz-Haber-Institut der Max-Planck-Gesellschaft,
Faradayweg 4-6, 14195 Berlin, Germany}

\author{Cynthia E. Heiner}
\affiliation{Fritz-Haber-Institut der Max-Planck-Gesellschaft,
Faradayweg 4-6, 14195 Berlin, Germany}

\author{Hendrick L. Bethlem}
\affiliation{Laser Centre Vrije Universiteit, De Boelelaan 1081,
1081 HV Amsterdam, The Netherlands}

\author{Andr{\'e} J.A. van Roij}
\affiliation{Institute for Molecules and Materials, Radboud
University Nijmegen, Heyendaalseweg 135, 6525 AJ Nijmegen, The
Netherlands}

\author{Gerard Meijer}
\affiliation{Fritz-Haber-Institut der Max-Planck-Gesellschaft,
Faradayweg 4-6, 14195 Berlin, Germany}

\date{\today}

\begin{abstract}
The level of control that one has over neutral molecules in beams dictates
their possible applications. Here we experimentally demonstrate that
state-selected, neutral molecules can be kept together in a few mm long
packet for a distance of over one mile. This is accomplished
in a circular arrangement of 40~straight electrostatic hexapoles through
which the molecules propagate over 1000~times. Up to 19~packets of
molecules have simultaneously been stored in this ring structure. This
brings the realization of a molecular low-energy collider within reach.
\end{abstract}


\maketitle
The important role that molecular beams have played ever since the
early days of quantum mechanics stems from the exquisite control
that one can exert over the internal and external degrees of freedom
of neutral molecules in these beams. Advances in the control over
molecular beams have gone hand in hand with new applications. In the
1930s, Rabi invented a method to unravel the quantum structure of
molecules, based on the controlled manipulation of the trajectories of
molecules on their way from an effusive source to the detector \cite{Rabi1939}.
When the deflection fields from these original experiments were replaced
by focusing elements, better control and higher densities were obtained
in the molecular beam experiments, leading to the invention of the
maser \cite{Gordon1955}. Improved control has also been obtained by
upgrading the sources. Molecular beams with kinetic energies above
one eV -- sufficient to study and overcome reaction barriers in selected
systems -- were obtained by gas dynamic acceleration, i.e., by seeding
small amounts of heavier molecules in light carrier gases \cite{Abuaf1967}.
The cooling of the translational, rotational and vibrational degrees of
freedom of molecules in these seeded mixed-gas expansions, has greatly
reduced the complexity in molecular spectra, and has thereby revolutionized
molecular spectroscopy \cite{Levy1981}.

During the last decade, various methods have been demonstrated to control
the forward velocity of molecules in beams. Decelerated beams of neutral
molecules have been used, for instance, to reduce the transit-time broadening in
high-resolution spectroscopy, to tune the collision energy in scattering studies,
and to enable accurate measurements of lifetimes of long-lived states \cite{Meerakker2008}.
Here, we report that the distance that these
decelerated molecular beams travel can be extended to over one mile. This
has been achieved by arranging a large number of straight electrostatic
hexapoles in a circle. The motion of a compact packet of state-selected
molecules through this circular arrangement is fully under control; the long
distance travelled testifies to the intrinsic stability of the molecular trajectories.

The motivation for constructing the circular arrangement of straight hexapoles
is its anticipated use as a low-energy collider; counter-propagating molecules
meet many times when they revolve around the ring. This makes it possible to
study low-energy (below $\approx$20~K) molecule-molecule collisions, probing the inter-molecular
interaction potential in great detail \cite{RefColdMol}. Molecular storage rings of various designs
have been theoretically analyzed \cite{Katz1997,Nishimura2003}, and
a compact electrostatic storage ring, obtained by bending a linear hexapole
focuser into the shape of a torus, has been experimentally demonstrated
\cite{Crompvoets2001}. In this prototype storage ring, the molecules were
transversely confined but longitudinally, i.e., along the circle, they were essentially
free. Therefore, an injected packet of molecules spread out along the flight
direction and eventually filled the entire ring. Collision studies can be
most sensitively performed when the neutral molecules revolve in bunches,
i.e., when they are both transversely and longitudinally confined. It has been recently
demonstrated that the latter can be achieved in a storage ring composed of
two half rings, separated by small gaps; packets of ammonia molecules were
kept together by changing the electric fields synchronous with the molecules'
passage through the gaps \cite{Heiner2007}.

In this prototype molecular synchrotron, the molecules could ultimately be kept
together for about hundred round trips, implying that the molecules passed
about 200~times through a gap, traveling a distance of about 80~meters
during a total storage time of about one second \cite{Heiner2009a}. At this time
the density of the stored packet was about two orders of magnitude below the
value at the time of injection. In collision studies involving counter-propagating packets
of molecules, losses in the number of stored molecules due to (in)elastic
collisions accumulate over the interaction events. Let us consider $n$ packets of
molecules revolving clockwise and $n$ packets revolving
anti-clockwise, all with the same speed; this would require a structure consisting out of
$n$ segments. After $m$ round trips, each packet of molecules would
have interacted $2\cdot n\cdot m$ times with counter-propagating packets. As the
summed intensity of the $n$ packets (revolving in either direction) can be recorded,
an additional factor $\sqrt{n}$ will be gained in sensitivity. By tuning the speed with
which the packets revolve, the total collision cross-section can be measured
as a function of collision energy. Therefore, to operate it as a low-energy collider,
the number of simultaneously stored packets -- and thus the number of ring
segments -- as well as the number of completed round trips need to be made
as large as possible.

Attempts to scale up the prototype molecular synchrotron to one consisting of
many segments met with a number of serious problems: (i) Curved electrodes
(as used in \cite{Heiner2007}) are difficult to machine to a high precision. This
causes the alignment of the ring to be rather poor, and thereby limits the lifetime of
the stored molecules. (ii) The original scheme to bunch the molecules involved
switching between four different voltage configurations every time a packet passed
through a gap \cite{Heiner2007}. For a ring consisting of many short segments the
load on the high voltage switches would be unacceptable. Moreover, (iii) we have
come to realize that this bunching scheme led to additional losses from the ring,
either due to instabilities of the molecular trajectories or due to non-adiabatic
transitions to non-trapped states. The molecular synchrotron presented here
overcomes all these limitations.

\begin{figure}[t]
\centering
\includegraphics[width=0.5\textwidth]{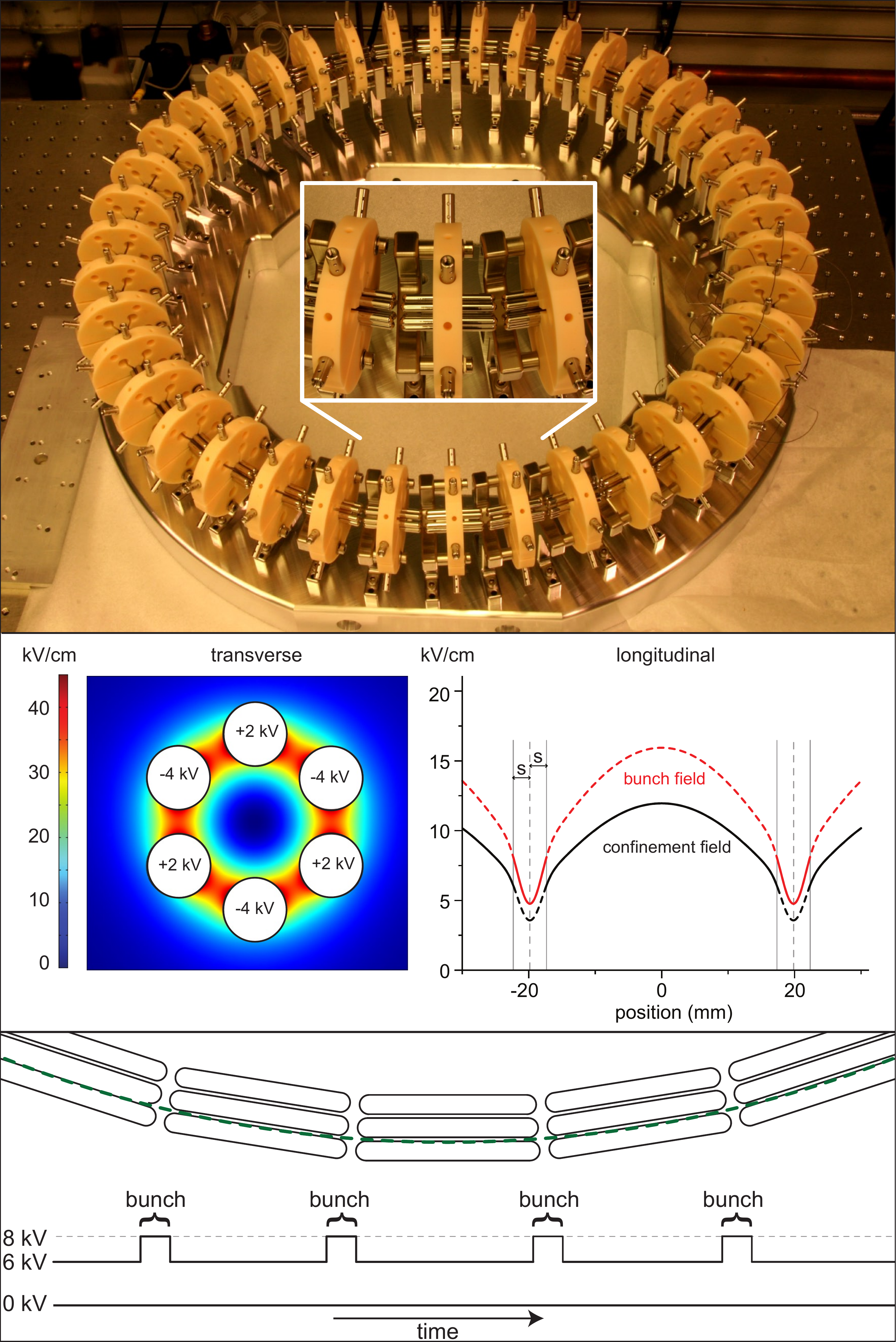}
\caption{(Color online) Upper panel: Photograph of the 0.5 m diameter molecular
synchrotron with a zoom-in of three hexapoles.
Middle panel, left: Voltage configuration used to transversely confine
neutral molecules, overlayed with a contour plot of the resulting
electric field strength. Right: Electric field strength
(both for confinement and bunching) as a function of position
along the equilibrium orbit.
Lower panel: Scheme of the time-dependence of the voltage difference
between adjacent hexapole electrodes during operation of
the synchrotron. By temporarily switching to a higher voltage-difference
whenever the packet passes through a gap, ammonia molecules with a
velocity spread of one~m/s are kept in a tight bunch while revolving.
The green dashed curve indicates the equilibrium orbit of ND$_3$
molecules with a forward velocity of 125~m/s.
\label{fig:BunchingScheme}}
\end{figure}

A hexapole torus can be approximated by many short straight
hexapole segments. From numerical simulations, we expect that the
number of straight hexapoles can be reduced to below 100 without
suffering a significant reduction in the transverse acceptance of
the ring. As a compromise between construction demands and the
expected performance, we have built a ring out of 40~straight
hexapoles. A photograph of this ring is shown in the upper panel
of Figure \ref{fig:BunchingScheme}, along with a zoom-in of three
of the hexapoles. The diameter of the ring is 500 mm.
Individual hexapoles are made by placing (nominally)
37.4~mm long electrodes, rounded off at each end, with a
diameter of 4~mm on the outside of a circle with a radius of
3.54~mm \cite{Anderson1997}. Each hexapole spans an angle of
9~degrees of the ring. A difference of 1.5~mm between the lengths
of the inner and outer electrodes is implemented to maintain a
constant gap of 2~mm between adjacent hexapoles.

To confine the molecules transversely, a voltage difference of 6~kV is applied
between adjacent hexapole rods, as shown in the middle panel of Figure
\ref{fig:BunchingScheme}. In this situation, the resulting electric field is cylindrically
symmetric, increasing quadratically with distance from the center. Molecules in
states that have a positive energy shift in the applied electric
field -- so-called low-field seekers -- will experience a force
towards the center of the hexapole; this force increases linearly
with the radial distance when the Stark shift is linear with the
electric field. In the ring, the molecules also experience a
centrifugal force that depends on their forward velocity, and
these forces will cancel at a certain position. In a perfect hexapole torus,
a deuterated ammonia (ND$_3$) molecule in the low-field-seeking component of the
$\mid$J,K$\rangle$=$\mid$1,1$\rangle$ state with
a forward velocity of 125~m/s would be displaced radially outwards
by 2.4~mm from the geometric center of the hexapole. In the present segmented
ring, it is still a sufficiently good approximation that the equilibrium orbit for
this molecule is a circle with a radius of 252.4 mm. Molecules
flying with the same forward velocity but with a different radial position or with
a nonzero radial velocity will oscillate around this equilibrium orbit with a transverse
(betatron) frequency of around 1~kHz, i.e., these molecules will
make 11.6~betatron oscillations per round trip.

Let us next look at the longitudinal motion of molecules on the equilibrium orbit.
As is seen in the middle panel of Figure \ref{fig:BunchingScheme}, the electric
field strength varies strongly along this orbit. This arises from the change in the distance
of the equilibrium orbit from the centerline of the straight hexapole segment and
from the presence of the gaps. The electric field in the middle of a gap drops to
about half of the value near the end of a hexapole segment. Therefore, molecules
in a low-field-seeking state will be accelerated in the fringe fields at the end of a
hexapole segment, and will be decelerated in the fringe fields at the next hexapole
segment. If the voltages are kept at a constant value, the deceleration will exactly cancel
the acceleration and the molecules will keep their original velocity. In order to confine
molecules in the longitudinal direction, we temporarily increase the positive voltages
from +2~kV to +4~kV every time the molecules pass through a gap. As schematically
indicated in Figure \ref{fig:BunchingScheme}, we increase the fields when a so-called
'synchronous' molecule is a distance $s$ before the middle of the gap and decrease
the fields when this molecule is the same distance $s$ after the middle of the gap.
As before, the accelerating and decelerating forces on the synchronous molecule
will be equal, and the velocity of the synchronous molecule will be the same after passing the
gap. By contrast, a molecule that is slightly behind the synchronous molecule will enter the
gap when the voltages are high, and it will leave the gap when the voltages are low. As a
consequence, this molecule will experience a net acceleration. Vice versa, a molecule that
is in front of the synchronous molecule will enter the gap when the voltages are low, and it
will leave the gap when the voltages are high. This molecule will experience a net deceleration.
This argument shows that molecules within a small position interval will experience a force
towards the synchronous molecule and will oscillate around it; the molecules are hence
trapped in a travelling potential well that revolves around the ring.
From measurements and simulations of this bunching scheme, we have determined the
longitudinal trap depth to be about 1~mK, and the trap (synchrotron) frequency to be on the
order of 35~Hz, i.e., the molecules make 0.4~synchrotron oscillation per round trip. Note that
in this bunching scheme the molecules experience a transverse confinement force at all times.

In our experiments, packets of deuterated ammonia ($^{14}$ND$_3$) molecules with a
forward velocity of around 125~m/s are tangentially injected into the 40-segment molecular
synchrotron. These packets are produced by Stark deceleration, followed by transverse
focusing and longitudinal cooling, of a pulsed beam of ammonia, as described in detail
elsewhere \cite{Heiner2006}. At the entrance of the synchrotron, each packet is several
mm long and contains about a million molecules. All of these molecules are in the
upper inversion doublet component of a single rotational level,
$\mid$J,K$\rangle$=$\mid$1,1$\rangle$, in the vibrational and electronic ground state.
At the injection point, four
hexapoles are connected (in two sets of two hexapoles) to two separate high-voltage
switches to allow for the injection of new packets of molecules without interfering with
the ones already stored. These segments are switched off during injection of a new packet,
and are switched to a small bias voltage during detection of the molecules. The remaining 36~hexapoles
are connected to three other high voltage switches.

\begin{figure}[tb] \centering
\includegraphics[width=0.5\textwidth]{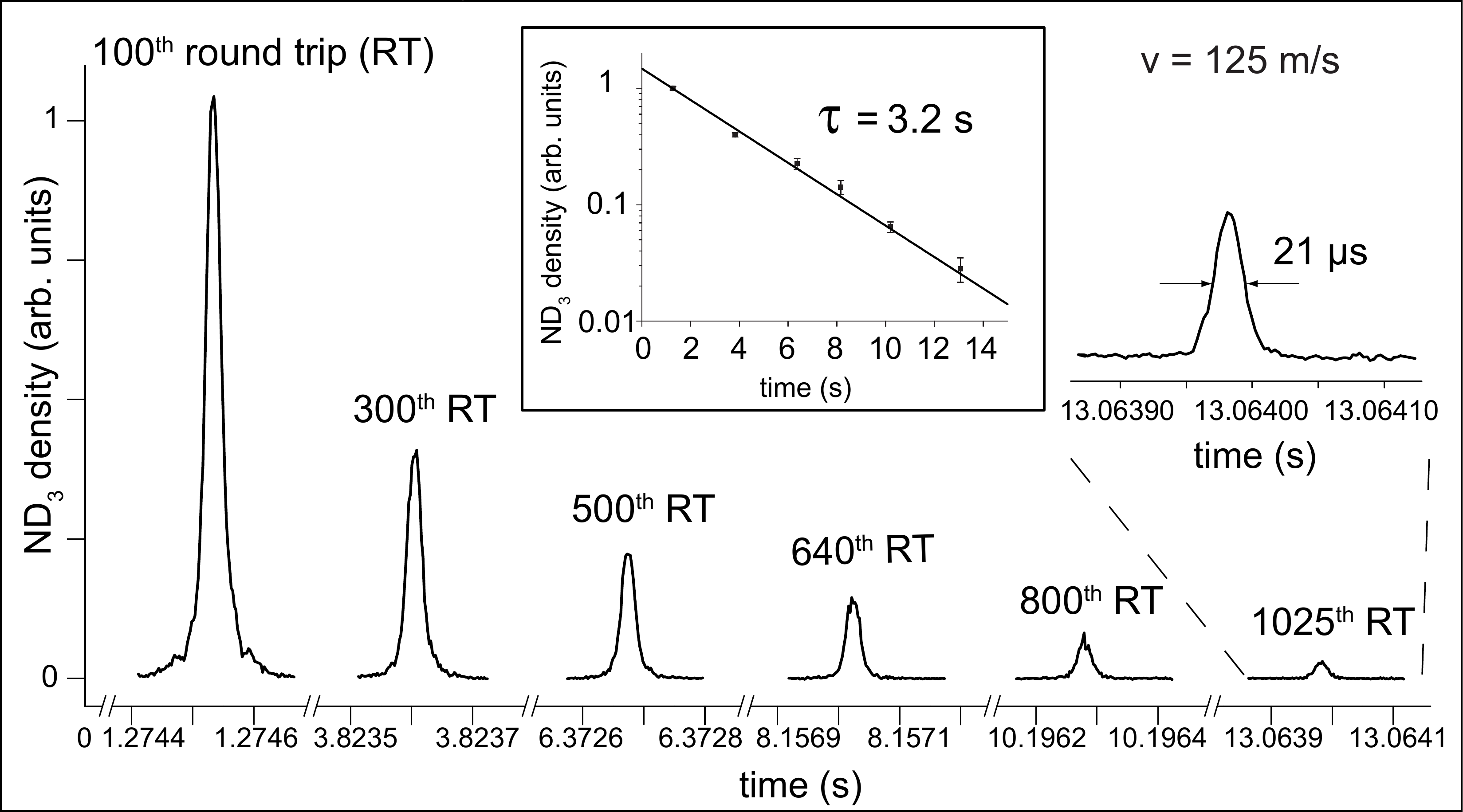}
\caption{Measurements of the density of ND$_3$ molecules as a function of time (in seconds)
for a selected number of round trips. The observed temporal width of 21~$\mu$s after
1025 round trips corresponds to a packet length of 2.6~mm after a flight distance of more
than a mile. The inset shows the exponential decay of the signal with time.
\label{fig:1milemeasurement}}
\end{figure}

Figure \ref{fig:1milemeasurement} shows the density of ammonia molecules as a function of
time after injection into the synchrotron. Thirteen packets are injected at a rate of about 10~Hz,
after which the loading is stopped. These packets then trail each other by a distance of three
hexapoles; the first and the last injected packet are four hexapoles apart. The summed signal
of the packets is recorded after a selected number of round trips, in time steps of 2.6~$\mu$s,
using ionization with a focused laser beam at the injection point. The time-of-flight profiles shown
are averaged over 16 of such measurement cycles. Even after 1025~round trips, i.e., after the
molecules have travelled further than a mile and have passed through a gap 41,000~times,
their signal can be clearly recognized; the temporal width of 21~$\mu$s corresponds to
a packet length of 2.6~mm.

The density of ammonia molecules is seen to exponentially decay with time at a rate of
0.31 $s^{-1}$. This is the lowest decay rate that has been observed for neutral ammonia
molecules in any trap to date. In all the early electrostatic trapping experiments,
$1/e$-lifetimes of only a small fraction of a second have been observed. These were probably
limited by non-adiabatic transitions to non-trappable states near the trap center, as was only
realized when substantially longer lifetimes of up to 1.9~seconds were measured in an
electrostatic trap with a non-zero electric field at the center \cite{Kirste2009}. With the present
confinement and bunching scheme, the molecules are never close to a zero electric field in the
ring either. In addition, we never change the direction -- but only the magnitude -- of the electric field
in the ring, effectively preventing the occurrence of non-adiabatic transitions \cite{Wall2010}.
A major contribution to the
observed loss-rate is optical pumping of the ammonia molecules out of the
$\mid$J,K$\rangle$=$\mid$1,1$\rangle$ level by blackbody radiation, calculated to occur
at a rate of 0.14 $s^{-1}$ in the room-temperature chamber \cite{Hoekstra2007}. The remaining
loss-rate of 0.17 $s^{-1}$ is well explained by collisions with background gas at the approximately
5$\cdot$10$^{-9}$ mbar pressure in the vacuum chamber.

\begin{figure}[htpb]
\centering
\includegraphics[width=0.5\textwidth]{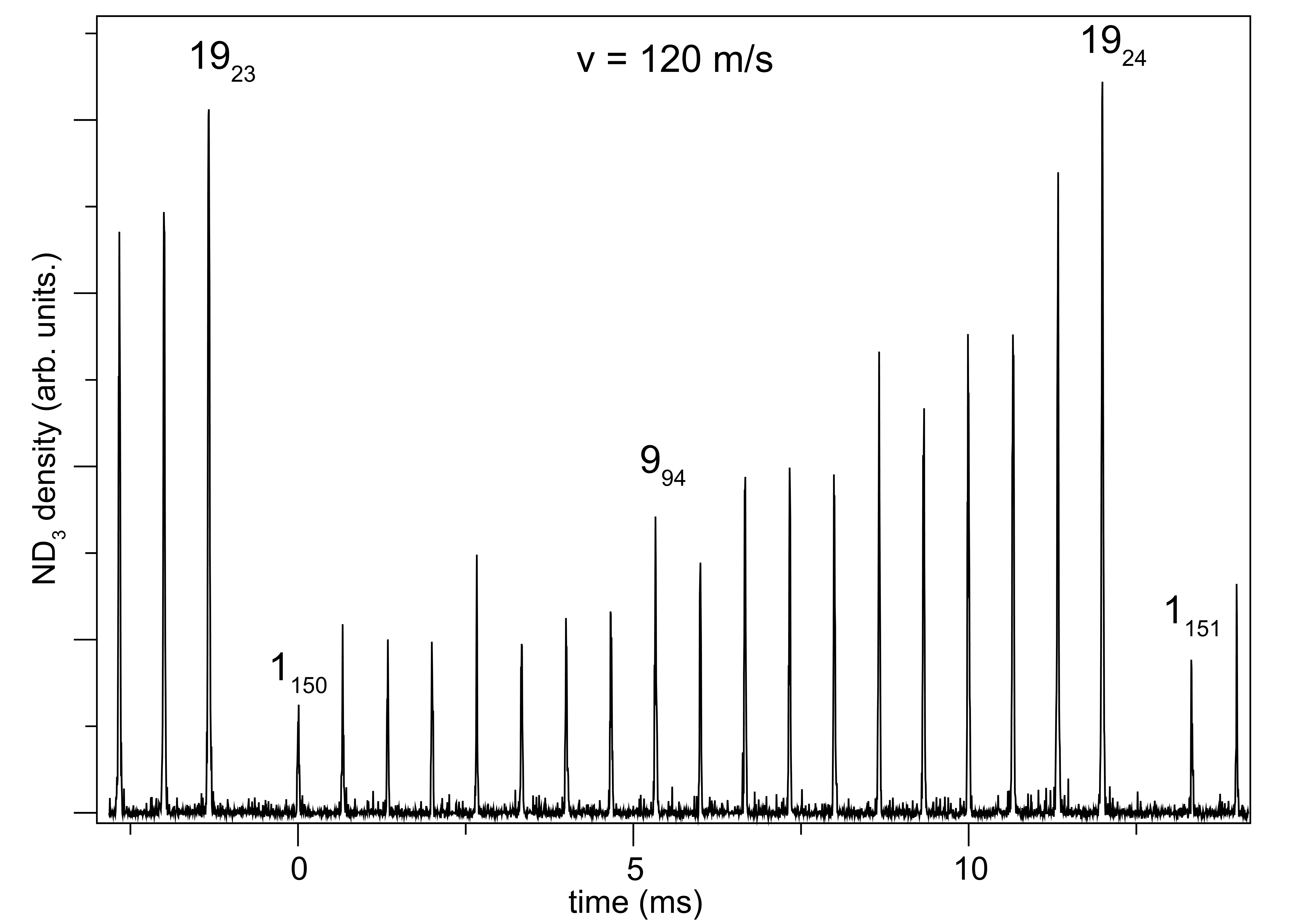}
\caption{Time-of-flight measurement showing 19~packets of ammonia molecules revolving
simultaneously in the molecular synchrotron. The horizontal axis shows the time
relative to the time at which the first injected packet has completed 150~round trips. Each packet
is labelled with two numbers: the main number labels the order of injection whereas the subscript
labels the number of completed round trips.
\label{fig:19packets}}
\end{figure}

Figure \ref{fig:19packets} shows a measurement of nineteen packets of ammonia molecules
with a forward velocity of 120~m/s being
injected at about 10~Hz into the ring. The horizontal axis shows the time
relative to the time at which the first injected packet has completed 150 round trips.
The last injected packet has been in the ring for 320~ms and has made 24~round trips
while the first injected packet has by then already been stored for more than 2 seconds.
The loading scheme is set up such that each packet completes
7~round trips plus the length of 2~hexapoles before the next packet is injected. In the
measurement the packets are seen to trail each other by 666~$\mu$s -- precisely the
time it takes them to fly through two hexapoles. In principle, it would be straightforward to
store 40~packets simultaneously in the present circular arrangement, but in our current
setup, we need to turn off at least two hexapoles to allow for new packets to enter the ring.

In conclusion, a synchrotron for neutral molecules consisting of 40~straight
hexapoles together with a simple bunching scheme has been presented. Up to
19~packets of ammonia molecules have simultaneously been stored in this structure.
The stored molecules can still be readily detected after they have completed a
distance of over one mile, corresponding to a storage time of over 13~seconds.
They have then passed -- confined in a compact packet -- 41,000 times through
a gap, which explicitly demonstrates the stability of their trajectories. These measurements epitomize
the level of control that can now be achieved over molecular beams and set the stage for
novel experiments to come. For collision studies, the sensitivity of this ring is
increased by more than three orders of magnitude over that of the prototype
molecular synchrotron \cite{Heiner2007}. We are currently setting up a second beamline to
inject counter-propagating packets into the ring.

\begin{acknowledgments}
We acknowledge the design of the electronics by G. Heyne, V. Platschkowski, and T. Vetter
and the software support by U. Hoppe. H.L.B. acknowledges financial support from NWO
via a VIDI-grant, and from the ERC via a Starting Grant.
\end{acknowledgments}

\end{document}